\documentclass{article}
\usepackage{spconf,amsmath,graphicx, amssymb}
\usepackage{subcaption}
\usepackage{bm}
\usepackage{multirow}
\usepackage{booktabs}
\usepackage{tikz}
\usepackage{float}
\newcommand{\RNum}[1]{\uppercase\expandafter{\romannumeral #1\relax}}

\title{One model to enhance them all: array geometry agnostic multi-channel personalized speech enhancement}
\name{\begin{tabular}{c}Hassan Taherian$^{1,2*}$, Sefik Emre Eskimez$^{1}$, Takuya Yoshioka$^{1}$, Huaming Wang$^{1}$, \\ 
Zhuo Chen$^{1}$ {\normalfont{and}} Xuedong Huang$^{1}$ \thanks{*Work done during internship at Microsoft Research.}\end{tabular}}
\address{$^1$Microsoft, Redmond, WA, USA~~~~~~~~~~~~
$^2$The Ohio State University, Columbus, OH, USA\\{\small \texttt{taherian.1@osu.edu, \{seeskime, tayoshio, huawang, zhuc,  xdh\}@microsoft.com}}}

\begin{document}
\ninept
\maketitle

\begin{abstract} 


With the recent surge of video conferencing tools usage, providing high-quality speech signals and accurate captions have become essential to conduct day-to-day business or connect with friends and families. Single-channel personalized speech enhancement (PSE) methods show promising results compared with the unconditional speech enhancement (SE) methods in these scenarios due to their ability to remove interfering speech in addition to the environmental noise. In this work, we leverage spatial information afforded by microphone arrays to improve such systems' performance further. We investigate the relative importance of speaker embeddings and spatial features. Moreover, we propose a new causal array-geometry-agnostic multi-channel PSE model, which can generate a high-quality enhanced signal from arbitrary microphone geometry. Experimental results show that the proposed geometry agnostic model outperforms the model trained on a specific microphone array geometry in both speech quality and automatic speech recognition accuracy. We also demonstrate the effectiveness of the proposed approach for unseen array geometries.

\end{abstract}

\section{Introduction}
\label{sec:intro}

Video conferencing has become essential in the emerging hybrid work era catalyzed by the COVID-19 pandemic. Most modern telecommunication services are equipped with a causal/real-time speech enhancement (SE) front-end to deliver high-quality speech audio in noisy environments. Recently,   ``personalized'' SE methods are emerging in the research field by utilizing an enrollment utterance of a target speaker as additional information to not only suppress the ambient noise and reverberation but also remove interfering speech \cite{scPSE, giri2021personalized}. With the ability to handle overlapped speech, personalized speech enhancement (PSE) models significantly improve the perceptual speech quality and the performance of downstream tasks such as automatic speech recognition (ASR)~\cite{scPSE}.

The SE performance can be improved further by using microphone arrays. With multiple microphones, spatial information can be extracted and combined with spectral information for obtaining better SE models \cite{wang2018all, chakrabarty2019time, tolooshams2020channel}. This paper extends the PSE~\cite{scPSE} to utilize the microphone arrays for environments where strong noise, reverberation, and an interfering speaker are present. We show that the combination of the enrolled speaker's embedding and the spatial features can significantly improve the SE performance. We also examine the impact of the speaker embedding and spatial features in challenging conditions where the target and interfering speakers have similar angles or distances to the array.

In addition, we introduce an array-geometry-agnostic PSE model that works regardless of the number of microphones and the array shape. This enables us to train the model for once and use it across multiple microphone array devices without additional operations such as adaptation or retraining, allowing the developed solution to scale significantly. Meanwhile, the geometry agnostic model also yields consistent improvements over fixed-geometry enhancement networks that are trained for matched array geometries. We also show the effectiveness of our proposed model for unseen array geometries and discuss its limitations.

\section{Related Work}

Several studies utilized speaker embeddings to extract the target speaker voice in speech separation tasks~\cite{wang2018voicefilter, wang2019speech, delcroix2020improving, Li2019direction}. In~\cite{delcroix2020improving}, spatial features from a binaural setup were included as an auxiliary cue to improve the separation performance for same-gender mixtures.~\cite{Li2019direction} incorporated a set of fixed beamformers and an attention network to select the dominant beam using the target speaker embedding. The speaker embeddings have also been employed for SE. A perceptually motivated PSE model with low complexity was proposed in~\cite{giri2021personalized}. \cite{scPSE} introduced two real-time PSE models and tested with reverberant target speech corrupted by both noise and interfering speech. We employ their personalized DCCRN (pDCCRN) model as our single-channel baseline system.

Along a related direction, multi-channel geometry invariant modeling was explored recently
~\cite{luoEND, wang2020seperation, zhang2021microphone, yemini2021scene}. In~\cite{luoEND}, the authors proposed a transform-average-concatenate (TAC) layer for multi-channel speech separation that was invariant to the order of the microphones. An inter-channel processing layer based on self-attention was proposed in~\cite{wang2020seperation}. The work by~\cite{yemini2021scene} used deep symmetric sets layers based on a Siamese network for speech dereverberation.~\cite{zhang2021microphone} proposed a model to process a variable number of microphone pairs for speech enhancement. 


\begin{figure*}[ht]
  
  \includegraphics[width=0.99\textwidth]{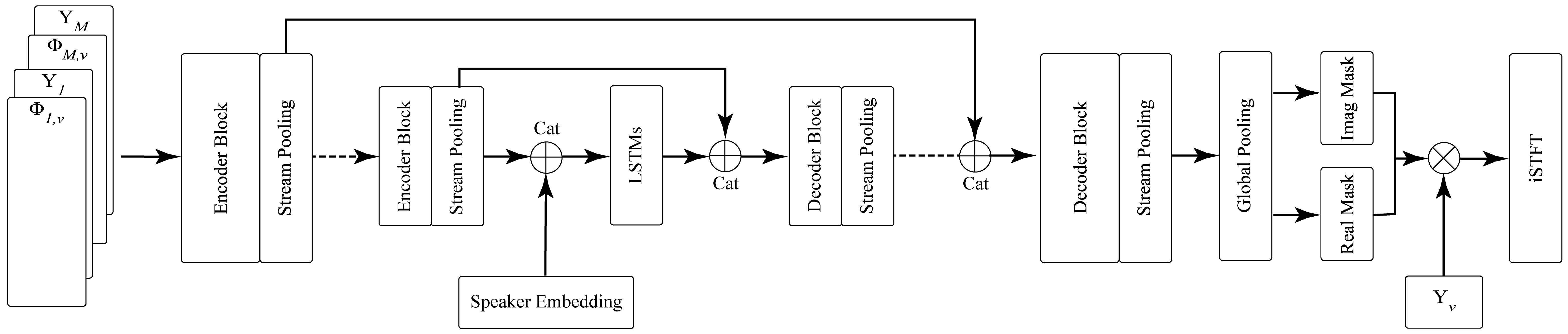}
  \centering
  \caption{Geometry agnostic multi-channel PSE model is shown. Skip connections link encoders to the corresponding decoders. 
  $Y_M$ and $\Phi_{M,v}$ represent STFT and IPD features of microphone~$M$, respectively. ‘Cat’ refers to concatenation and symbol $\otimes$ is point-wise complex multiplication.}
  \label{fig_diagram}
  \end{figure*}


Compared to the aforementioned models, our proposed geometry agnostic model has a more straightforward design. To process inputs with variable dimensions (i.e., microphones), we introduce stream pooling layers for convolutional neural networks that are only based on the averaging and concatenation of the feature maps. 
Furthermore, our experiments focus on real-time processing scenarios and are conducted on an extensive list of array geometries, which have not been previously explored.




\section{System Description}

\subsection{Baseline: Single-channel PSE}
\vspace{-.4em}

The single-channel PSE model, based on which we build our multi-channel models, performs complex spectral mapping using pDCCRN~\cite{scPSE, Hu2020DCCRNDC}. pDCCRN has a U-Net architecture with encoder and decoder blocks and two complex LSTM layers in between. Each block contains complex 2-D convolutional layers followed by complex batch normalization. Complex layers are formed by two separate real-valued layers that operate on the real and imaginary parts of the layer input~\cite{dcunet}.
The pDCCRN uses a d-vector and the mixture signal $Y$ in the short-time Fourier transform (STFT) domain. The real and imaginary parts of the mixture signal are concatenated to form the input feature $x \in \mathbb{R}^{C \times 2F \times T }, C=1$, where $C$, $F$ and $T$ are the number of channels for convolution, frequency bins, and time frames. We replicate the d-vector through time dimension, concatenate it with the encoder's output, and feed the concatenated vector into the first LSTM layer.
The model outputs a complex ratio mask~\cite{crm} which is multiplied with the input mixture to estimate the clean speech. We use 6 encoder layers with $C = [16, 32, 64, 128 ,128, 128]$, kernel size of $5 \times 2$ and stride of $2 \times 1$.  The model is trained with a power-law compressed phase-aware mean-squared error (MSE) loss function~\cite{Eskimez2021HumanLA}. We emphasize that all operations are causal and operate in real-time. For details about our d-vector extractor, we refer the reader to~\cite{zhou2021resnext}.

\begin{table*}[ht]
  
  \caption{Evaluation results for fixed geometry multi-channel PSE models with a 7-channel circular array with a radius of $4.25$ cm are shown. `MC PSE' refers to multi-channel PSE. We show WER(\%), SDR (dB), STOI(\%). The best values are highlighted with bold font.}
 \vspace{-.5em}
   \centering
 \resizebox{\textwidth}{!}{
    \begin{tabular}{{l ccc c ccc c ccc c ccc }}
    \toprule
      &  \multicolumn{3}{c}{\textbf{A}} && \multicolumn{3}{c}{\textbf{B}} &&\multicolumn{3}{c}{\textbf{B (similar angle)}} && \multicolumn{3}{c}{\textbf{B (similar distance)}} \\
    \cmidrule{2-4} \cmidrule{6-8} \cmidrule{10-12} \cmidrule{14-16}
     & WER & SDR & STOI && WER & SDR & STOI && WER & SDR & STOI && WER & SDR & STOI     \\  
    \midrule
     Noisy mixture  &	22.38 &	5.65 &	73.54 &&	41.07 &	2.74 &	69.07 &&39.75	&2.63	&68.41	&&40.97	&2.64	&68.58 \\  
    \midrule

 Single-channel PSE &27.65	&10.79	&84.07	&&38.09	&8.55	&79.30 &&37.32	&8.60	&79.09	&&39.20	&8.07	&78.32 \\
    \midrule
    
    MC PSE (STFT) &20.73	&12.22	&88.52	&&24.91	&10.14	&85.04 &&28.35	&\textbf{9.56}	&83.59	&&25.85	&9.78	&84.48 \\
   \midrule
   
   MC PSE (IPD) &\textbf{19.88}	&\textbf{12.23}	&\textbf{88.78}	&&\textbf{22.42}	&\textbf{10.44}	&\textbf{86.38} &&\textbf{27.26}	&9.33	&\textbf{83.79}	&&\textbf{23.24}	&\textbf{9.93}	&\textbf{85.44}\\
   \midrule
   
   \hspace{3mm} --w/o d-vector &20.27	&11.82	&88.11	&&25.80	&9.40	&84.37&& 30.00	&8.53	&82.23	&&29.17	&8.19	&81.65\\
   
    \bottomrule

    \end{tabular}
  }
  \label{tab:7ch_feature}
\end{table*}

\subsection{Multi-channel PSE for Fixed Microphone Arrays} %
\vspace{-.4em}
We employ two approaches to extend the single-channel PSE model to $M$ microphones with a fixed array geometry. In the first approach, real and imaginary parts of all microphones STFT are stacked in the channel dimension ($C$) to create the input $ x \in \mathbb{R}^{M \times 2F \times T } $, which is fed to the PSE model. With this simple extension, the model can implicitly learn the spectral and spatial information \cite{chakrabarty2019time, wang2021multi}. 

In the second approach, we explicitly extract the spatial information. As with \cite{delcroix2020improving}, we use inter-channel phase difference (IPD) as the spatial features. The IPD for the microphone pair $(i,j)$ is defined as $\Phi_{i,j}  = \angle (Y_i / Y_j)  $.
The IPD features are calculated between the first microphone and each of the other $M-1$ microphones. For each pair, we  concatenate the cosine and sine values of the IPD features. Finally, we stack all IPD features, as well as the real and imaginary parts of the first microphone's STFT to form the input feature $x \in \mathbb{R}^{ M \times 2F \times T }$.

For both approaches, as with the single-channel case, the model estimates the complex-valued masks which would recover the clean signal when applied to the first microphone. 

\begin{figure}[t]
  
  \includegraphics[width=0.45\textwidth]{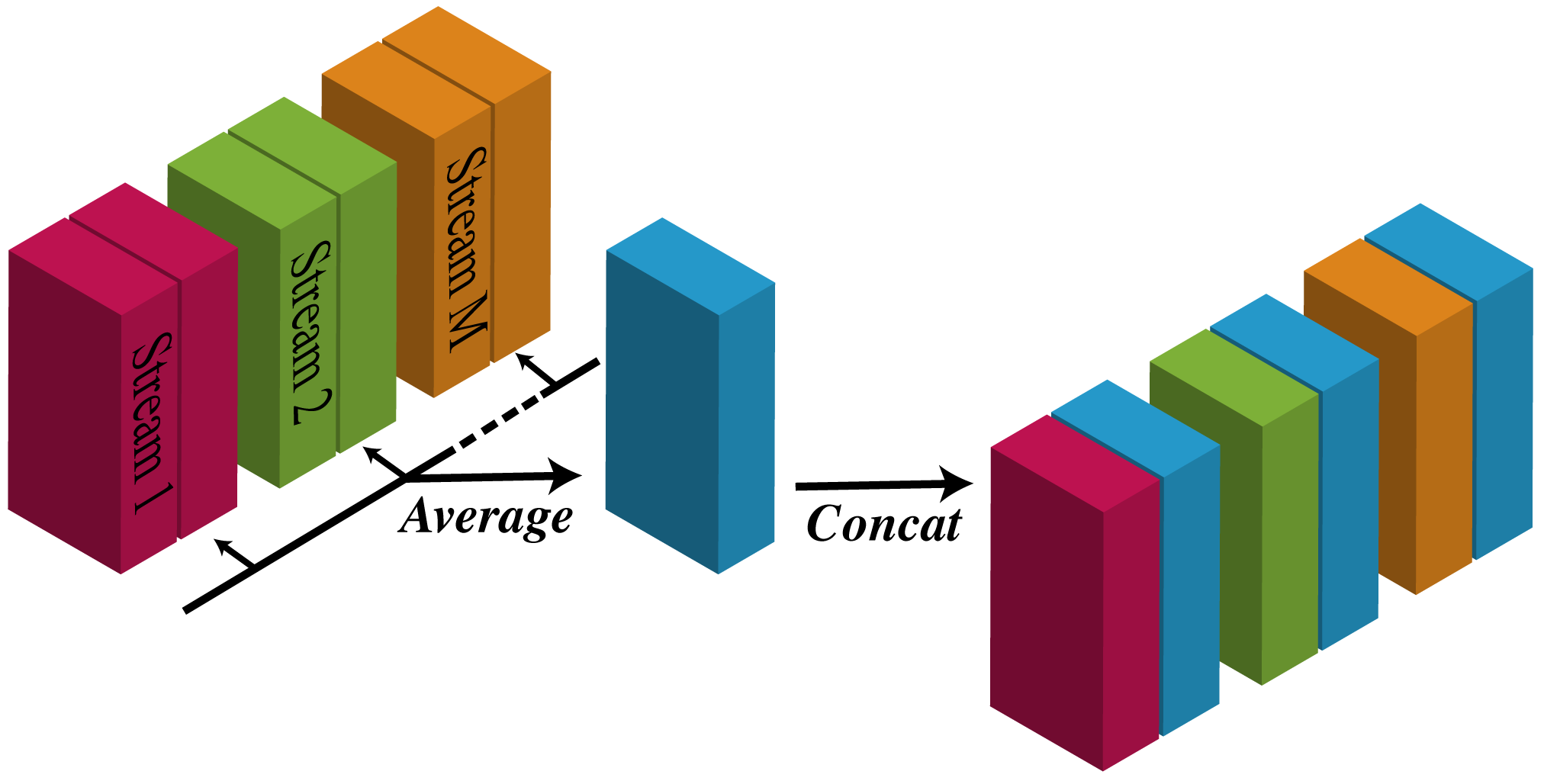}
  \centering
  \caption{Schematic diagram of stream pooling layer is shown. Each cube represents a stream (microphone) tensor.}
  \label{fig_steam_pooling}
  \end{figure}

\subsection{Geometry Agnostic Modeling}
\vspace{-.4em}
The proposed multi-channel geometry-agnostic PSE model can be described as follows. We first create a virtual microphone signal $Y_v$ by simply taking the average of all the input microphones:

\begin{equation}
  Y_v = \frac{1}{M}  \sum_{n = 1}^{M} Y_i.  
  \label{virtual_mic}
  \end{equation}

\noindent Then, we extract the IPD features for each microphone with respect to the virtual microphone: 
$\Phi_{i,v}  = \angle  ( Y_i /Y_v)$.
 We also normalize the IPD features using the unbiased exponentially weighted moving average \cite{chiley2019online} to increase the robustness of the model to arbitrary array geometries. Compared with the fixed geometry models, we observed that the IPD normalization was critical for the geometry agnostic models since the microphone arrangement could be different during training and testing.  

In the geometry agnostic model, we introduce an additional dimension to all layers that contain specific stream (microphone) information and append it as the first dimension of the input tensor of each layer. Thus, 
the model input $ x \in \mathbb{R}^{M \times  C \times 2F \times T } $  contains STFT and IPD features of each stream in the channel dimension, i.e., $C=2$. 
We refer to the first dimension as a stream. 

Figure~\ref{fig_diagram} illustrates the geometry agnostic model architecture. The model is applicable to any number of microphones, and the output is invariant to the permutation of the microphones. 
Given the input tensor described above, 
we process each stream information in parallel by using pDCCRN. To utilize the spatio-temporal patterns exhibited in the input multi-channel audio, we include a stream pooling layer after every encoder and decoder blocks of pDCCRN. In these layers, we split the channel dimension into two parts: one is unique to each stream; the other is used to aggregate information across the streams. Each cross-stream convolution channel is averaged across the streams and then appended to the stream-specific channels of each stream. A diagram of the stream pooling layer is shown in Fig.~\ref{fig_steam_pooling}. At the output layer of pDCCRN, we use a global-pooling layer to average across all the streams and channels to estimate complex masks. The estimated masks are applied to the virtual microphone STFT, $Y_v$.

\section{Experimental Setup}

We evaluated our multi-channel PSE models by using simulated data that covered various scenarios. We generated room impulse responses (RIRs) for a 7-channel circular microphone array with radius $r=4.25$ cm (see Table~\ref{tab:seenGmt}.\RNum{4})  based on the image method with T60 in the range of 0.15 to 0.6 seconds. The microphone array was located in the room center. Target and interfering speakers were positioned randomly around the array within [0.5, 2.5] meters with the assumption of the target speaker being always closer to the array.

For the training and validation datasets, we simulated 2000 and 50 hours of audio, respectively,  based on the clean speech data from the DNS challenge dataset \cite{reddy2021interspeech}. In both datasets, 60\% of utterances contained the target and interfering speakers with a signal-to-interference ratio (SIR) between 0 to 10 dB. The mixed audio was further corrupted by simulated isotropic noises and directional noises from the Audioset and Freesound datasets \cite{audioset,freesound} with a signal-to-noise ratio (SNR) in the range of [0, 15] dB. The sampling rate for all utterances was 16 kHz. We trained our geometry agnostic model with the 7-channel circular array and 3 other geometries derived from it: 4-channel triangular, 4-channel rectangular, and 6-channel circular arrays. 

We created two 10-hour test datasets called A and B. Dataset A contained utterances mixed only with ambient noise and reverberation. In contrast, dataset B contained utterances mixed with the ambient noise, reverberation, and interfering speech. Clean utterances were selected from internal conversational speech recordings with high neural network-based mean opinion score (MOS) values~\cite{pmos}. The SIR and SNR ranges were the same as in the training dataset. We convolved the test utterances with RIRs from 8 different geometries. Four geometries were the same as the ones used for the training dataset. The other four geometries were unseen during the training and included a 3-channel triangular array with $r=4.25$ cm, a 5-channel circular array with $r=3$ cm, a 3-channel linear array with 6 cm length, and an 8-channel circular array with $r=10$ cm (as used in the AMI corpus \cite{amicorpus}).

We evaluated the enhanced speech signal based on the word error rate (WER) and two signal quality metrics, i.e., signal-to-distortion ratio (SDR) and short-time objective intelligibility (STOI). We followed the setup described in \cite{Eskimez2021HumanLA} for the ASR evaluation.
We used two baselines for comparison with our geometry agnostic model. For each array geometry that was used for training, we trained a fixed geometry model based on the IPD features without normalization. The other baseline was based on processing each microphone independently with a single-channel PSE model followed by averaging the enhanced signals. Although this approach is computationally expensive, it is an acceptable alternative for unknown array geometries. Note that we also tried to use MVDR beamforming followed by the single-channel PSE; however, the results were worse, probably due to signal distortions caused by the real-time beamforming. Also, since we assumed that we did not have any knowledge about the microphone array geometries, it would be difficult, if not impossible, to do beamforming in real-time.

\section{Results and Discussions}

\begin{table*}   
  \caption{Performance comparison of fixed geometry and geometry agnostic multi-channel PSE models for seen microphone arrays is shown.}
 \vspace{-.5em}
  \centering
  \begin{tabular}{{llccc|ccc|ccc|ccc}}
    \toprule
    &&  \multicolumn{3}{c}{\includegraphics[scale=0.38]{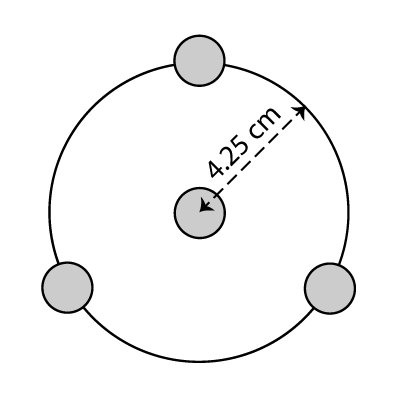}} &  \multicolumn{3}{c}{\includegraphics[scale=0.38]{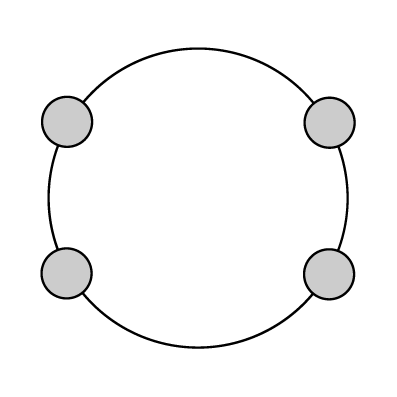}}
    &  \multicolumn{3}{c}{\includegraphics[scale=0.38]{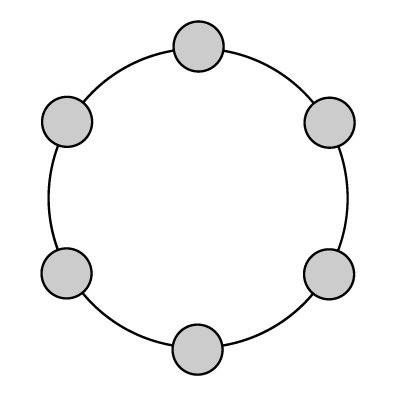}}
    &  \multicolumn{3}{c}{\includegraphics[scale=0.38]{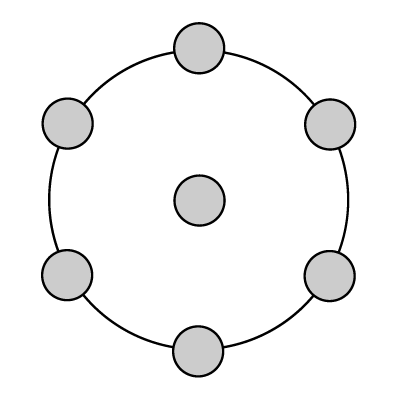}}\\

    &&  \multicolumn{3}{c}{\RNum{1}. Triangular (4ch)} &  \multicolumn{3}{c}{\RNum{2}. Rectangular (4ch)}     &  \multicolumn{3}{c}{\RNum{3}. Circular (6ch)}     &  \multicolumn{3}{c}{\RNum{4}. Circular (7ch)}\\

    \midrule 
    && WER & SDR & STOI & WER & SDR & STOI & WER & SDR & STOI & WER & SDR & STOI \\
    \midrule
    \multirow{3}{*}{\large{A}}
    & Signal Averaging &		25.30&	10.75&	84.89 &	25.80	&10.50	&84.34 &	24.93	&10.60	&84.69  &24.86	&10.73	&84.92 \\
    & Fixed Geometry &	    21.19&	11.50&	86.99 &	22.84	&11.15	&86.15 &	20.85	&11.48	&87.39  &	19.88	&12.23	&88.78 \\
    &Geometry Agnostic &	  \textbf{20.75}&	\textbf{12.02}&	\textbf{88.01} &\textbf{21.41}	&\textbf{11.79}	&\textbf{87.34} &  \textbf{19.99}	&\textbf{12.08}	&\textbf{88.41}  &\textbf{19.65}	&\textbf{12.24}	&\textbf{88.86} \\
    \midrule
    \multirow{3}{*}{\large{B}}
    & Signal Averaging &	34.72	&8.62	 &80.54     &	35.25&	8.44	&80.00 &	34.26&	8.53&	  80.45 &	34.06&	8.62	&80.68 \\
    & Fixed Geometry &	  24.95	&9.72	 &84.55     &	25.42&	9.64	&84.20 &	23.20&	9.96&	  85.36 &	22.42&	10.44	&86.38 \\
    &Geometry Agnostic &	\textbf{22.85} &\textbf{10.62} &\textbf{86.67}     &	\textbf{24.06}&	\textbf{10.36}	&\textbf{85.81} &	\textbf{21.78}&	\textbf{10.76}&	\textbf{87.34} &	\textbf{21.35}&	\textbf{10.90}	&\textbf{87.76} \\
    \bottomrule

  \end{tabular}
  \label{tab:seenGmt}  
\end{table*}

Table~\ref{tab:7ch_feature} shows the experimental results of different PSE models trained with the 7-channel circular array. Both STFT- and IPD-based multi-channel PSE models substantially outperformed the single-channel PSE model in all scenarios. As with previous studies \cite{Eskimez2021HumanLA, sato2021should}, we observed that the single-channel PSE  introduced processing artifacts that yielded worse WER scores compared to the unprocessed noisy mixture for dataset A (i.e., no interfering speech). By contrast, the multi-channel PSE models improved the speech recognition performance.

With regard to the comparison between the two spatial features for the multi-channel PSE, the model trained with the IPD features performed consistently better than the model based on the stacked STFTs. We also trained a multi-channel PSE model based on the IPD features without using d-vectors. The results show that spatial information was helpful regardless of the presence of d-vectors.

To further investigate the effect of the spatial features, we created two versions of test dataset B. In the first version, speakers were randomly positioned such that the difference of their angles with respect to the array was less than 5 degrees while their distance difference was more than 1 meter. In the second version, the angle difference of the speakers was more than 45 degrees, while the distance difference was less than 10 cm. We observed that the performance gap between the single-channel and multi-channel models was smaller with the similar speaker angle setting than with the other settings. 
This shows the usefulness of the speakers' directions for discriminating overlapping voices. 
Also, when the two speakers were at similar distances, using d-vectors substantially improved the performance of the multi-channel PSE.

\begin{table}[H] 
  \caption{Geometry agnostic multi-channel PSE results for unseen array geometries are shown.}
  \vspace{-.5em}
  \centering
 \resizebox{0.48\textwidth}{!}{
  \renewcommand{\arraystretch}{1.25}
    \begin{tabular}{{ccl ccc }}
    \toprule
    
     &&& WER & SDR& STOI \\  
    \midrule
    \multirow{5}{*}{\includegraphics[scale=0.38]{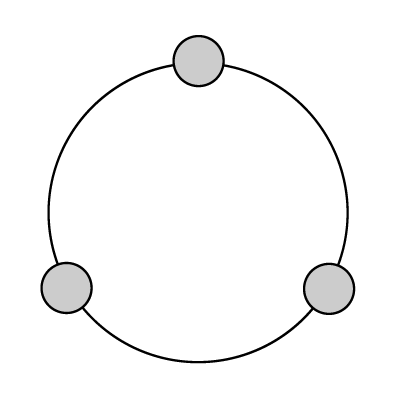}}  
     & \multirow{3}{*}{\textbf{A}} 
     &  Signal Averaging        &25.71	&10.52	&84.50	\\  
     && Fixed Geometry &	23.29	&11.22	&85.90	\\
     &&Geometry Agnostic &\textbf{22.10}	&\textbf{11.65}	&\textbf{86.96}\\
     \cmidrule{3-6}
     & \multirow{3}{*}{\textbf{B}} 
     & Signal Averaging        &34.91&	8.45	&80.14	\\  
     && Fixed Geometry &	25.62&	9.84	&84.23	\\
     Triangular (3ch)&&Geometry Agnostic &\textbf{24.53}&	\textbf{10.27}	&\textbf{85.51}\\ 

     \midrule
     \addlinespace[3pt]
     \multirow{3}{*}{\includegraphics[scale=0.38]{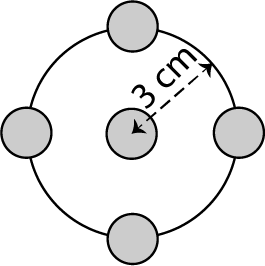}}  

      & \multirow{3}{*}{\textbf{A}} 
      &  Signal Averaging        &24.00	&11.41	&85.77	\\ [3pt]  
      &&Geometry Agnostic &\textbf{20.22}	&\textbf{12.51}	&\textbf{88.45}\\ [3pt]
      \cmidrule{3-6} 
      & \multirow{3}{*}{\textbf{B}} 
      & Signal Averaging        &34.51&	8.77	&80.73	\\ [3pt]
      Circular (5ch)&&Geometry Agnostic &\textbf{22.34}&	\textbf{10.90}	&\textbf{87.16}\\
      \midrule

      \multirow{4}{*}{\includegraphics[scale=0.38]{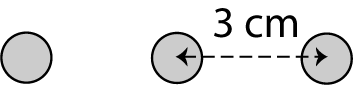}}  

      & \multirow{2}{*}{\textbf{A}} 
      &  Signal Averaging        &24.61	&11.37	&85.49	\\ 
      &&Geometry Agnostic &\textbf{23.99}	&\textbf{11.79}	&\textbf{86.07}\\
      \cmidrule{3-6} 
      & \multirow{2}{*}{\textbf{B}} 
      & Signal Averaging        &35.11	&8.73	&80.36	\\
      Linear (3ch)&&Geometry Agnostic &\textbf{28.82}	&\textbf{9.75}	&\textbf{83.51}\\
      \midrule
      \multirow{5}{*}{\includegraphics[scale=0.38]{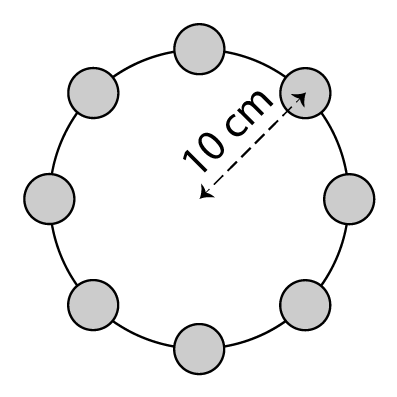}}  
       & \multirow{3}{*}{\textbf{A}} 
       &  Signal Averaging        &\textbf{22.28}	&8.93	&\textbf{84.19}	\\  
       
       &&Geometry Agnostic &25.99	&\textbf{9.33}	&82.27\\
       && \hspace{3mm} --w/o IPD Norm &	30.99	&8.48	&78.87	\\
       \cmidrule{3-6}
       & \multirow{3}{*}{\textbf{B}} 
       & Signal Averaging        &32.82&	7.03	&79.62	\\  
       
       &&Geometry Agnostic &\textbf{29.30}&	\textbf{8.11}	&\textbf{80.65}\\
       Circular (8ch) &&\hspace{3mm} --w/o IPD Norm &	35.41&	7.36	&77.01	\\
   
    \bottomrule

    \end{tabular}
  }
  \label{tab:unseenGmt}
\end{table}

Table~\ref{tab:seenGmt} shows the geometry agnostic multi-channel PSE results for the arrays seen during the training. Our proposed geometry agnostic model outperformed all the fixed geometry models trained for the corresponding array geometries in both test datasets. This result suggests that not only can our approach make the model independent of the microphone array geometries, but also it can model the relationship between multiple-channel feature streams more efficiently than simply concatenating them. 
Without requiring to change the model architecture for individual arrays, a single model can be shared between multiple arrays with different shapes and the number of microphones. Also, training with different geometries has an augmentation effect which improves the robustness compared with fixed geometry training.

Table~\ref{tab:unseenGmt} shows the geometry agnostic model results for the microphone arrays that were not used during the training. We observed that the geometry agnostic model still outperformed the fixed geometry model for the 3-channel triangular array, which has fewer microphones than the arrays included in the training. For the 5-channel circular array, which has a different microphone arrangement, the geometry agnostic model performed very well, achieving the performance comparable to the seen array geometries in Table~\ref{tab:seenGmt}. Regarding the 3-channel linear array, the geometry agnostic model showed consistent improvements over the average of enhanced single-channel signals despite not being exposed to the front-back ambiguity of the linear arrays during the training.

For the 8-channel circular array with $r=10$ cm, the geometry agnostic model improved the performance compared with the average of the enhanced signals to a smaller extent, and the results for dataset A were worse in terms of WER and STOI. We speculate the spatial aliasing was the reason for the relatively poor performance of the 8-channel circular array. A large inter-microphone distance leads to spatial aliasing, and this can introduce unseen patterns for the IPD features. For example, spatial aliasing occurs in the array geometries where its inter-microphone distance is longer than 4.28 cm with 16 kHz sampling rate~\cite{wolfel2009distant}. We observed that if the model was trained without IPD normalization, the performance degraded significantly, suggesting the spatial aliasing problem can be mitigated by IPD normalization.
\vspace{-.5em}
\section{Conclusions}
\vspace{-.5em}
In this work, we utilized spatial features along with speaker embeddings for personalized speech enhancement and showed their combination significantly improved the performance for both ASR and signal quality. Furthermore, we proposed a new architecture and introduced the stream pooling layer to perform multi-channel PSE with any number and arrangement of microphones in a way where the output is invariant to the microphone order. Our proposed model consistently outperformed the geometry-dependent models. 
Future challenges include mitigating the spatial aliasing problem. 

 
\bibliographystyle{IEEEtran}
{\small\bibliography{refs}}

\end{document}